\documentclass[onecolumn,noshowpacs,nofootinbib,10pt]{revtex4-1}

\usepackage{float,xcolor,upgreek,yfonts,tikz}
\usepackage{color,array,ctable}
\usepackage{graphicx,color}
\usepackage{graphicx,float}\usepackage[all]{xy}
\usepackage{amsmath,upgreek}

  \newcolumntype{?}{!{\vrule width 1pt}}
   \usepackage{caption}
 \usepackage{booktabs,makecell}
\usepackage{subcaption}
\usepackage{amssymb}

\usepackage{alphabeta}
\usepackage{dcolumn}
\usepackage{textgreek}
\usepackage{multirow} 
\usepackage{units}
\usepackage{overpic}

\newcommand{\beq}{\begin{eqnarray}}
\newcommand{\eeq}{\end{eqnarray}}
\newcommand{\be}{\begin{equation}}
\newcommand{\ee}{\end{equation}}

\newcommand{\bea}{\begin{eqnarray}}
\newcommand{\eea}{\end{eqnarray}}

\newcommand{\ba}{\begin{eqnarray}}
\newcommand{\ea}{\end{eqnarray}}
\usepackage[colorlinks,hyperindex,unicode]{hyperref}
\definecolor{green1}{RGB}{0,128,0} 
\hypersetup{hidelinks,backref=true,pagebackref=true,hyperindex=true,colorlinks=true,breaklinks=true,urlcolor= blue}
\hypersetup{%
  colorlinks = true,
  linkcolor  = blue,
  citecolor = cyan,
}
\usepackage{bookmark,textgreek}
\usepackage{hyperref,color,xcolor}
\hypersetup{hidelinks,hyperindex=true,colorlinks=true,breaklinks=true,urlcolor= blue}
\hypersetup{%
  colorlinks = true,
  linkcolor  = blue
}
\newcommand\orcidroldao{{\href{https://orcid.org/0000-0003-3978-532X}{\orcidicon}}}
\newcommand{\orcidicon}{%
	\begin{tikzpicture}
	\draw[lime, fill=lime] (0,0)
		circle [radius=0.16]
		node[white] {{\fontfamily{qag}\selectfont \tiny ID}};
	\draw[white, fill=white] (-0.0625,0.095)
		circle [radius=0.007];
	\end{tikzpicture}	\hspace{-2mm}
}

\newcommand{\JJ}{\mathbin{\raisebox{0.25ex}{$\footnotesize
                       \rm\vphantom{I}%
                       \_\hskip -0.25em\_%
                       \vrule width 0.6pt$}}}           

\newcommand{\JJBB}{\mathop{\JJ \,}\displaylimits_{\scalebox{0.65}{$\emph{g}$}}} 
 
\newcommand{\JJgg}{\mathop{\JJ \,}\displaylimits_g} 
\newcommand{\JJee}{\mathop{\JJ \,}\displaylimits_\eta} 

\newcommand{\ii}{\imath}
\newcommand{\jj}{\jmath}
\newcommand{\kk}{k}

\newcommand{\cl}{C\ell}     

\def\Re{\mathop{\rm Re}}

\newcommand{\ut}[1]{{\setbox0=\hbox{$#1$}\mathsurround=0pt
       \rlap{\raisebox{-0.8\dp0}{\raisebox{-0.8ex}
       {\kern -0.15ex\hbox{$\tiny\sim$}\kern 0.15ex}}}#1}}
\newcommand{\uti}[1]{{\setbox0=\hbox{$#1$}\mathsurround=0pt
       \rlap{\raisebox{-0.8\dp0}{\raisebox{-0.8ex}
       {\kern -0.3ex\hbox{$\tiny\sim$}\kern 0.3ex}}}#1}}

\newdimen\arrayruleHwidth                 
     \makeatletter
     \def\Hline{\noalign{\ifnum0=`}\fi\hrule \@height \arrayruleHwidth
         \futurelet \@tempa\@xhline}
     \makeatother

\newcommand{\bege}{\begin{equation}}
\newcommand{\enge}{\end{equation}}
\newcommand{\benu}{\begin{enumerate}}
\newcommand{\enu}{\end{enumerate}}

\newcommand{\bbbbox}{\mathop{\Box\kern -5pt\raisebox{.8pt}{$|$}}}

\newcommand{\clt}{\color{cyan}}
\newcommand{\cle}{C\ell_{1,3}}

\newcommand{\CC}{\mathbb{C}}

\def\beq{\begin{eqnarray}}
\def\eeq{\end{eqnarray}}

\def\0{\mbox{\boldmath$\displaystyle\mathbf{0}$}}

\newcommand{\pu}{\underset{\!\!\!\!\!{\scalebox{0.65}{$\emph{g}$}}}}

\newcommand{\clb}{\cle^\mathit{g}}

\begin{document}

\title{
Emergent spinor fields from exotic spin structures}

\author{J. M. Hoff da Silva}
\email{julio.hoff@unesp.br} 
\affiliation{DFI, Universidade Estadual Paulista, Unesp, Guaratinguet\'{a}, 12516-410, Brazil}

\author{R. da Rocha\orcidroldao\!\!}
\email{roldao.rocha@ufabc.edu.br}
\affiliation{Center of Mathematics,  Federal University of ABC, 09210-580, Santo Andr\'e, Brazil.}

\begin{abstract}{}
The classification of emergent spinor fields according to modified bilinear covariants is scrutinized, in spacetimes with nontrivial topology, which induce inequivalent spin structures. Extended Clifford algebras, constructed by equipping the underlying spacetime with an extended bilinear form with additional terms coming from the nontrivial topology, naturally yield emergent extended algebraic spinor fields and their subsequent extended bilinear covariants, which are contrasted to the classical spinor classification. An unexpected duality between the standard and the exotic spinor field classes is therefore established, showing that a complementary fusion process among the spinor field classes sets in, when extended Clifford bundles are addressed in multiply-connected spacetimes. 
\end{abstract}
\maketitle
\section{Introduction}

The development of Clifford algebras has been fundamental to several aspects of mathematics and physics. One can go back to Cartan, who described Clifford algebras as matrix algebras and, in addition, proposed and showed the periodicity mod 8 theorem. Besides, he introduced the concept of the spinor, as well as the spinor as the square root of geometry, including the pure spinor ramification. The term spinor was coined by Ehrenfest in quantum mechanics, although the intrinsic concept of a spinor had been employed long before Ehrenfest used it. The concept of spinor represented a landmark that has made Clifford algebras play a prominent role in physics, concomitantly with Pauli introducing the concept of spin to physics. Dirac exhibited a first-order partial differential equation in the relativistic formulation of quantum mechanics, based upon a Clifford algebra and their representation in Minkowski spacetime, describing the electron by a spinor field. One year after, Lanczos provided a quaternionic description of the Dirac equation, with Juvet and Sauter replacing column 2-spinors with square matrix spinors with the second column of zero entries. This last formulation was the prototype of considering spinors as elements of a minimal left ideal in a Clifford algebra, which has been formally implemented by Riesz in any dimension and metric signature \cite{oxford}. Clifford algebras are extremely useful when constructing spinor fields and associated Dirac operators, as well as periodicity and index theorems endowing a rich algebraic, geometric, and topological derived structure \cite{chevalley}. 

Spinor fields, according to the classical definition, carry irreducible representations of the Spin group, constructed upon the twisted Clifford-Lipschitz group of a spacetime, endowed with a symmetric bilinear form. This construction is possible when one regards the Stiefel--Whitney classes as topological invariants of a real vector bundle describing the obstructions to establish independent sets of sections of the vector bundle. Therefore the bundle admits a spin structure if and only if both the first and second Stiefel--Whitney classes do vanish \cite{Bonora:2009ta}. Spinor fields are the seed of bilinear covariants, which are homogeneous multivector fields involving quadratic arrangements of a given spinor, its dual, and the Clifford algebraic structure of gamma matrices. When taking into account the usual four-dimensional Minkowski spacetime structure, spinor fields were classified by Lounesto, from the point of view of the bilinear covariants \cite{lou2}, which was generalized to include the Hopf algebra structure and new classes of algebraic spinor fields in Ref. \cite{Ablamowicz:2014rpa}. The most general classification of spinor fields, in any spacetime dimension and metric signature, was obtained in Ref. \cite{Bonora:2014dfa}. Ref. \cite{Bonora:2014dfa} also obtained new classes of spinor fields as solutions of the equations of motion generated from supergravity coupled to gauge fields, with a dynamical mechanism of compactification. Thereafter, the Lorentzian case was added to the Riemannian case in Ref. \cite{Bonora:2015ppa}, whereas the second quantization procedure of these generalized spinor fields was introduced in Ref. \cite{Bonora:2017oyb}. 
The bilinear covariants are correlated by the Fierz--Pauli--Kofink equations. Other relevant spinor field classifications were engendered in Refs. \cite{Cavalcanti:2014wia}. Several relevant aspects of spinor field classification were investigated in Refs. \cite{fabb1}. Flag-pole and flag-dipole have been recently found in fermionic sectors of AdS/CFT \cite{Meert:2018qzk}. 
Other spinor representations were studied in Refs. \cite{Beghetto:2017nmb}, with other proposals to construct the bilinear covariants for flag-pole spinors \cite{HoffdaSilva:2016ffx}.

A relevant development was paved in Ref. \cite{DaRocha:2020oju}, wherein exotic spinor fields that emerge from multiply-connected manifolds were shown to trigger off additional terms composing the Dirac operator, governed by the cohomology group, for spacetimes of arbitrary dimension and metric signature. The corresponding heat kernel coefficients as well as the spectrum of the exotic Dirac operators were shown to yield the geometric invariants to carry unequivocal corrections and physical signatures of the nontrivial topology \cite{exotic}. There are several relevant applications of modeling physical systems on non-trivial topologies in spacetime. It includes thermodynamical systems, quantum field theory, gravity, superconductivity, and, more generally, condensed matter systems. The quantum theory governing various types of quantum fields, which propagate on multiply-connected manifolds, was reported in Refs. \cite{f2,Isham:1978ec,field,Dantas:2015mfi}. The very existence of a (non-trivial) line bundle on $M$ emulates the construction of scalar fields in multiply connected manifolds. In fact, manifolds that  are not simply-connected,  underlie the existence of inequivalent spin structures. When the hypotheses underlying Geroch's theorem are satisfied, spin structures exist in a manifold $M$. They are not unique and many inequivalent spin structures can exist \cite{Geroch:1968zm,Geroch:1970uv}. The set of real line bundles on $M$ and the set of inequivalent spin structures are well-known to be categorized by objects in the cohomology group $H^1(M,\mathbb{Z}_2)$, comprising the homomorphism group of the fundamental group, $\pi_1(M)$ into $\mathbb{Z}_2$. It is worth emphasizing that simply-connected manifolds have a vanishing fundamental group, and have a single spin structure. Diverse patches of the local double coverings can be made equivalent to distinct inequivalent spin structures, whose leading use in quantum gravity and string theory was addressed in  Refs. \cite{Atiyah:1973ad,Hawking:1977ab,Hawking:1978pog,Seiberg:1986by,Back:1978zf}. When quantum field theories are approached in spacetimes with non-trivial topology, 
among the eventual several inequivalent spin structures, one must select one of them to construct the Dirac operator entering the Dirac equation. Ref. \cite{Avis:1979de} discusses the construction of quantum field theories in multiply-connected spacetimes, stating that a possible choice consists of taking an appropriately weighted average of all existing spin structures, analogously to the weighted average sum of topological sectors when instantons are taken into account and also similarly to summing Feynman diagrams in quantum field theory.  In this way, Feynman's  path integral formulation must comprise multiply-connected manifolds in application to quantum theories of gravity \cite{Hawking:1978pog,Christensen:1978tw}. Multiply-connected manifolds can also be employed to study superconductors. Inequivalent spin structures can explain the Bardeen--Cooper--Schrieffer pairing in superconductors, and spinor fields in multiply-connected manifolds account for the Josephson effect in superconductors  \cite{petry,valk}. Investigating inequivalent spin structures on multiply-connected manifolds also shed new light on the instanton compactification and also in the problem of confinement \cite{Luscher:1977cw,Sasaki:1978pv,Schechter:1977qg}. 
Finite-temperature quantum field theories were scrutinized on multiply-connected manifolds \cite{Unwin:1979dt,Unwin:1980bn}.
Approaching the vacuum polarization in spinor electrodynamics, the choice of different inequivalent spin structures yields the equations of motion to generate distinct physical effects. 
The most natural choice consists of selecting an ulterior spin structure, different from the standard one, to compute the vacuum polarization upon the photon propagation. In fact, the standard spin structure yields a non-causal effect, whereas inequivalent spin structures produce the physically expected causal effect \cite{Ford:1979ds}. 
Exotic spin structures correspond to a vacuum state having energy lower than the vacuum associated with the standard spin structure   \cite{Ford:1979pr}. Many-body quantum systems were also studied and reported on manifolds with non-trivial topology, including the Hubbard model \cite{Boada:2014xfa}.

Within the context of Cartan's formalism and exotic spinors, Ref. \cite{JHEP} proposed a way to encompass a nontrivial topology-motivated term leading to a seminal effect into the spacetime metric. Such a term, by turn, is responsible for physical effects, such as the arising of quasinormal-like modes for a scalar field. Here we investigate the impact of such deformation in the spacetime metric on the spinor classification. We first construct the concept of extended spinor and related (extended) Clifford algebra by encompassing the nontrivial effects via the Clifford product. Then we move to study Lounesto's classification of the extended spinors. An interesting duality between both (extended and usual) classification schemes is shown. A complementary fusion process among the spinor field classes is implemented, for extended Clifford bundles in multiply-connected spacetimes.
  
The goal of this paper is, then, to provide a complete classification of extended algebraic spinor fields in the context of extended Clifford algebras and associate them with bilinear covariants, constructing an important duality between the spinor fields and their extended counterparts. The paper is organized as follows: in Section~\ref{w2}, standard concepts such as Clifford algebras, bilinear covariants, and associated spinor field classification are revisited. In Section~\ref{scle}, the correspondence between classical and algebraic spinor fields is obtained. These review sections are relevant to the extension we will accomplish further. In Section \ref{nv}, we pinpoint the essential aspects of merging Cartan's spinors to the exotic spinor framework. In Section~\ref{scle1}, extended algebraic spinor fields and their important properties are introduced and investigated. In Section~\ref{scle2}, the spinor field classification, according to their bilinear covariants, is described in the extended Clifford algebraic formalism. The duality between spinor fields and extended spinor fields is provided and summarized in Table~\ref{tata}. In Section~\ref{scle3} the conclusions are addressed. 

\section{Bilinear Covariants, Fierz identities, and Fierz aggregates}
\label{w2}

Hereon $M$ denotes a manifold of arbitrary finite dimension. Given $\uppsi$ as an element of the exterior bundle $\Upomega(M) =\oplus _{k=0}^{n}\Upomega^{k}(M)$, the 
 {reversion} is an anti-automorphism given by $\tilde{\uppsi}=(-1)^{[k/2]}\uppsi$, for $[k]$ denoting the integer part of $k$. Equipping $M$ by a metric  
${\eta}:\Upomega(M)\times \Upomega(M)\rightarrow \mathbb{R}$, it can be extended to $\Upomega(M)$, to define contractions. In fact, given any $\upkappa ={{a}}^{1}\wedge \cdots \wedge {{a}}^{i}\in\sec\Upomega^i(M)$ and $\upvarphi ={{c}}^{1}\wedge \cdots \wedge {{c}}^{l}\in\sec\Upomega^l(M)$, the extension ${\eta}_\Upomega: \Upomega(M)\times\Upomega(M)\to\mathbb{R}$ is such that ${\eta}_\Upomega(\upkappa ,\upvarphi )=\det ({\eta}(a^i,c^j))$, for $a^i,c^j$ denoting 1-forms. For any differential form $\upxi \in \sec\Upomega (M)$, the {left contraction} emulates the inner product for the exterior bundle as ${\eta}_\Upomega(\upkappa \lrcorner \upvarphi, \upxi)={\eta}_\Upomega(\upvarphi ,\tilde{\upkappa}\wedge \upxi )$, whereas the {right contraction} reads  
${\eta}_\Upomega(\upkappa \llcorner \upvarphi ,\upxi)={\eta}_\Upomega(\upvarphi ,\upkappa \wedge \tilde{\upxi})$. The Clifford product between ${v}\in \sec\Upomega^{1*}(M)$ and $\upkappa \in \sec\Upomega (M)$ can be constructed upon the contraction and the exterior product, and denoted by juxtaposition, as  
${{v}}\upkappa ={{v}}\wedge \upkappa +{{v}}\lrcorner \upkappa$. Hence the 2-tuple $(\Upomega (M),{\eta})$ is the Grassmann bundle that, when equipped with the additional Clifford structure, is the Clifford bundle $\cl (M,{\eta})$ or 
$\cl _{p,q}$, where the tangent bundle reads $TM\simeq \mathbb{R}^{p,q}$, for $p+q=\dim(M)$, where $(p,q)$ is the signature of $\eta$. 
When one considers the Minkowski spacetime corresponding to $p=1$ and $q=3$, the usual bilinear covariants can be introduced. The set $\{{\scalebox{0.8}{$\textsc{E}$}}^{\mu }\}$ denotes a basis of the space of sections in the coframe bundle $\mathbb{P}_{\mathrm{SO}_{1,3}^{e}}(M)$. In this way, the so-called (classical) spinor fields are well-known objects that carry the $\rho={\scalebox{0.8}{D}}^{(1/2,0)}\oplus {\scalebox{0.8}{D}}^{(0,1/2)}$ irreducible representation of the component of the Spin group connected to the identity, $\mathrm{Spin}_{1,3}^{e}$, in $\mathbb{C}^{4}$. Classical spinor fields are vector bundle sections of $\mathbb{P}_{\mathrm{Spin}_{1,3}^{e}}(M)\times _{\rho }\mathbb{C}^{4}$ \cite{Rodrigues:2005yz}. 

Given a spinor field $\uppsi \in \sec \mathbb{P}_{\mathrm{Spin}_{1,3}^{e}}(M)\times _{\rho }\mathbb{C}^{4}$, and its standard dual $\bar\uppsi=\uppsi ^{\dagger }\upgamma _{0}$, bilinear covariants are defined as homogeneous $k$-covector fields, sections of the exterior 
bundle~$\Upomega(M)$ \cite{lou2}:
\begin{subequations}
\beq
\!\!\!\!\!\!\!\!\!\!\!\upsigma &=&\bar\uppsi\uppsi\in\sec\Upomega^{0}(M),\label{fierz0}\\ \label{fierz1}\!\!\!\!\!\!\!\!\!\!\!\mathbf{{\scalebox{0.9}{$\textbf{J}$}}}&=&J_{\mu }{\scalebox{0.8}{$\textsc{E}$}} ^{\mu }=\bar\uppsi\upgamma _{\mu }\uppsi {\scalebox{0.8}{$\textsc{E}$}} ^{\mu }\in\sec\Upomega^{1}(M),\\ \label{fierz2}
\!\!\!\!\!\!\!\!\!\!\!\!\!\!\mathbf{{\scalebox{0.9}{$\textbf{S}$}}}\!&\!=\!&\!S_{\mu \nu }{\scalebox{0.8}{$\textsc{E}$}} ^{\mu }\wedge {\scalebox{0.8}{$\textsc{E}$}} ^{\nu }\!=\!\frac{1}{2}\bar\uppsi i\upgamma _{[\mu}\upgamma _{ \nu] }\uppsi {\scalebox{0.8}{$\textsc{E}$}} ^{\mu }\wedge {\scalebox{0.8}{$\textsc{E}$}} ^{\nu }\in\sec\Upomega^{2}(M)\\
\!\!\!\!\!\!\!\!\!\!\!\mathbf{{\scalebox{0.9}{$\textbf{K}$}}}&=&K_{\mu }{\scalebox{0.8}{$\textsc{E}$}} ^{\mu }=\bar\uppsi i\upgamma _{0}\upgamma _{1}\upgamma _{2}\upgamma _{3}\upgamma _{\mu }\uppsi {\scalebox{0.8}{$\textsc{E}$}} ^{\mu }\in\sec\Upomega^{3}(M),\label{fierz3}\\ \!\!\!\!\!\!\!\!\!\!\!\upomega&=&-\bar\uppsi\upgamma _{0}\upgamma _{1}\upgamma _{2}\upgamma _{3}\uppsi\in\sec\Upomega^{4}(M).  
\label{fierz}
\eeq
\end{subequations}
The set $\{\mathbb{I}_{4},\upgamma _{\mu },\upgamma _{\mu }\upgamma _{\nu },\upgamma _{\mu }\upgamma _{\nu }\upgamma _{\rho },\upgamma _{0}\upgamma _{1}\upgamma _{2}\upgamma _{3}\}$ is a basis of the Clifford bundle of Minkowski spacetime, whose multivector fields have representations on the space of $4\times 4$ gamma matrices, satisfying the Clifford relation $\upgamma _{\mu }\upgamma _{\nu }+\upgamma _{\nu }\upgamma _{\mu }=2\eta _{\mu \nu }\mathbb{I}_{4}$. 

\begin{figure}[H]
\centering
	\includegraphics[width=8cm]{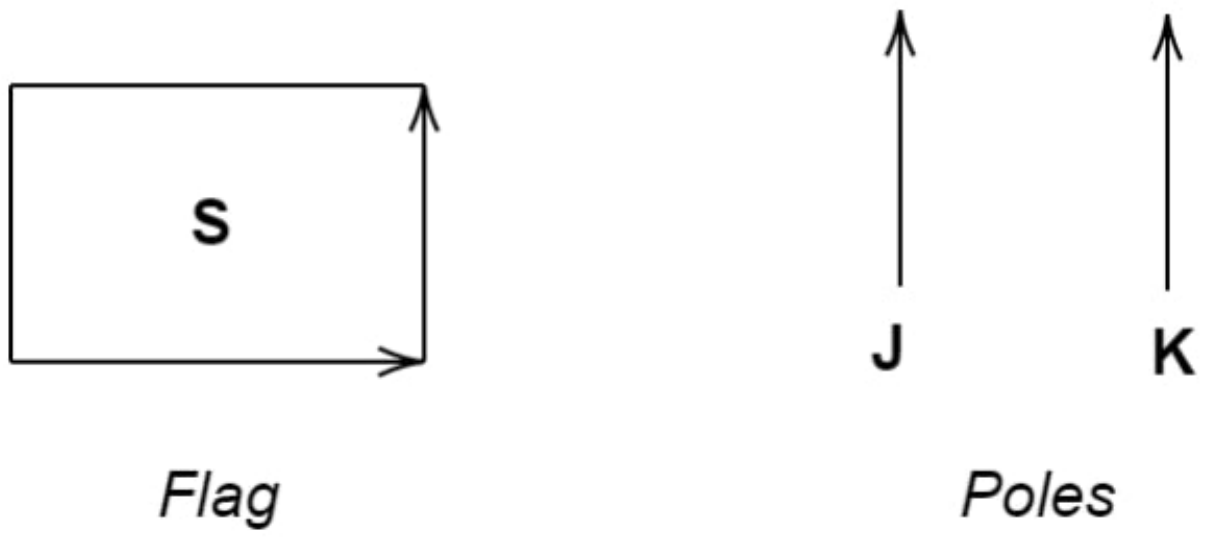}\vspace*{-1.7cm}
\caption{Pictorial representation of the bilinear covariants $\mathbf{J}$, $\mathbf{S}$, $\mathbf{K}$.}
\label{f1}
\end{figure}

\begin{figure}[H]
\centering
	\includegraphics[width=12cm]{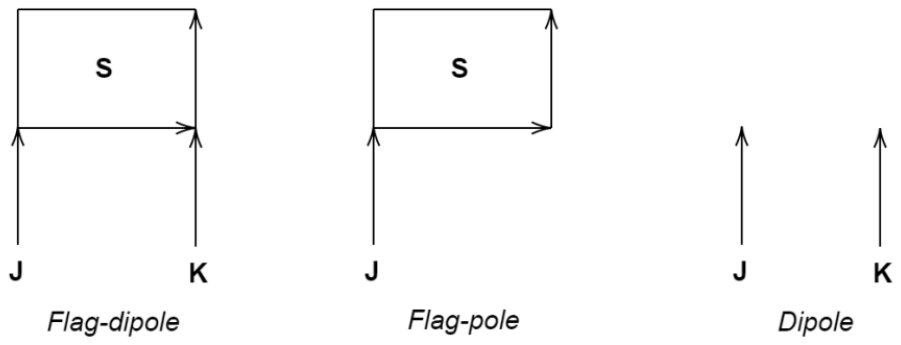}\vspace*{-6cm}
\caption{Pictorial representation of singular classes of spinor fields.}
\label{f2}
\end{figure}

Classical spinor fields can be classified into six disjoint groupings, given by the following spinor field classes~\cite{lou2}, where in the first three classes it is implicit that $\mathbf{{\scalebox{0.9}{$\textbf{J}$}}}, \mathbf{{\scalebox{0.9}{$\textbf{K}$}}}, \mathbf{{\scalebox{0.9}{$\textbf{S}$}}}\neq 0$:
\begin{itemize}
\item[1)] $\upsigma\neq0,\;\;\; \upomega\neq0\;\;\; \mathbf{{\scalebox{0.9}{$\textbf{K}$}}}\neq0\;\;\;\mathbf{{\scalebox{0.9}{$\textbf{S}$}}}\neq 0$,

\item[2)] $\upsigma\neq0,\;\;\; \upomega= 0\;\;\; \mathbf{{\scalebox{0.9}{$\textbf{K}$}}}\neq0\;\;\;\mathbf{{\scalebox{0.9}{$\textbf{S}$}}}\neq 0$,\label{dirac1}

\item[3)] $\upsigma= 0, \;\;\;\upomega\neq0\;\;\; \mathbf{{\scalebox{0.9}{$\textbf{K}$}}}\neq0\;\;\;\mathbf{{\scalebox{0.9}{$\textbf{S}$}}}\neq 0$,\label{dirac2}

\item[4)] $\upsigma= 0 = \upomega, \;\;\;\mathbf{{\scalebox{0.9}{$\textbf{K}$}}}\neq0,\;\;\; \mathbf{{\scalebox{0.9}{$\textbf{S}$}}}\neq0$,\label{tipo4}

\item[5)] $\upsigma= 0 = \upomega, \;\;\;\mathbf{{\scalebox{0.9}{$\textbf{K}$}}}= 0, \;\;\;\mathbf{{\scalebox{0.9}{$\textbf{S}$}}}\neq0$,\label{type-(5)1}

\item[6)] $\upsigma= 0 = \upomega, \;\;\; \mathbf{{\scalebox{0.9}{$\textbf{K}$}}}\neq0, \;\;\; \mathbf{{\scalebox{0.9}{$\textbf{S}$}}} = 0$.
\end{itemize}
\noindent 
In all these six classes, $\mathbf{{\scalebox{0.9}{$\textbf{J}$}}}\neq0$ is assumed, since its time component $J^0=\uppsi^\dagger\uppsi=\|\uppsi\|^2\neq0$ is the density of current probability. 
Besides, $\mathbf{{\scalebox{0.9}{$\textbf{J}$}}}$ is the current density, the bivector 
$\mathbf{{\scalebox{0.9}{$\textbf{S}$}}}$ provides the distribution of intrinsic angular momentum, and $\mathbf{{\scalebox{0.9}{$\textbf{K}$}}}$ regards the chiral current density. Spinor fields in the first three classes are called regular spinor fields, whereas the last three singular spinor classes are respectively called {flag-dipole}, {flag-pole}, and {dipole spinor fields}, due to a pictorial model by Penrose \cite{pen}. By the definitions (\ref{fierz1}) -- (\ref{fierz3}), the current density $\mathbf{{\scalebox{0.9}{$\textbf{J}$}}}$ is a 1-form and the chiral current density $\mathbf{{\scalebox{0.9}{$\textbf{K}$}}}$ is the dual of a 1-form, each one thus identified to a pole. On the other hand, $\mathbf{{\scalebox{0.9}{$\textbf{S}$}}}$ is a 2-form, having the geometric portray of a 2-dimensional plaquette in Minkowski spacetime, and can be identified as a flag. The Penrose interpretation can be depicted in Figs. \ref{f1} and \ref{f2}.\vspace*{-1.7cm}
Now, since type-4 spinor fields have $\mathbf{{\scalebox{0.9}{$\textbf{K}$}}}\neq0,$ $ \mathbf{{\scalebox{0.9}{$\textbf{S}$}}}\neq0$, and $\mathbf{{\scalebox{0.9}{$\textbf{J}$}}}\neq0,$ then there are two poles and a flag. Type-4 spinor fields and hence called flag-dipole spinors. Also, since type-5 spinor fields have $\mathbf{{\scalebox{0.9}{$\textbf{K}$}}}=0,$ $ \mathbf{{\scalebox{0.9}{$\textbf{S}$}}}\neq0$, and $\mathbf{{\scalebox{0.9}{$\textbf{J}$}}}\neq0,$ there are two poles and a flag, but one of these poles is null. Therefore, type-5 spinor fields and hence called flag-pole spinors. Since type-6 spinor fields have $\mathbf{{\scalebox{0.9}{$\textbf{K}$}}}\neq0,$ $ \mathbf{{\scalebox{0.9}{$\textbf{S}$}}}=0$, and $\mathbf{{\scalebox{0.9}{$\textbf{J}$}}}\neq0,$ there are two poles and a null flag. Hence type-6 spinor fields are named dipole spinor fields.

\vspace*{-7cm}
Several aspects of the flag-poles and flag-dipoles were proposed in Refs \cite{Fabbri:2020elt,5Fabbri:2020elt}, including applications in quantum gravity \cite{dasi}.

 The bilinear covariants satisfy the so-called Fierz identities \cite{Cra,hol}:
\begin{eqnarray}
\mathbf{{\scalebox{0.9}{$\textbf{J}$}}}^{2}&=&\upomega^{2}+\upsigma^{2},\qquad\mathbf{{\scalebox{0.9}{$\textbf{K}$}}}^{2}=-\mathbf{{\scalebox{0.9}{$\textbf{J}$}}}^{2},\qquad\mathbf{{\scalebox{0.9}{$\textbf{J}$}}}\llcorner\mathbf{{\scalebox{0.9}{$\textbf{K}$}}}=0,\nonumber\\\mathbf{{\scalebox{0.9}{$\textbf{J}$}}}\wedge\mathbf{{\scalebox{0.9}{$\textbf{K}$}}}%
&=&-(\upomega+\upsigma\upgamma_0\upgamma_{1}\upgamma_{2}\upgamma_3)\mathbf{{\scalebox{0.9}{$\textbf{S}$}}}. \label{fi}
\end{eqnarray}
\noindent 
One can also consider the Fierz aggregate as the multivector field \beq
Z=\upsigma+\mathbf{{\scalebox{0.9}{$\textbf{J}$}}}+i\mathbf{{\scalebox{0.9}{$\textbf{S}$}}}+i\mathbf{{\scalebox{0.9}{$\textbf{K}$}}}\upgamma_0\upgamma_{1}\upgamma_{2}\upgamma_3+\upomega\upgamma_0\upgamma_{1}\upgamma_{2}\upgamma_3,\eeq whose homogeneous components are the bilinear covariants of a given spinor field. Whenever the Fierz aggregate $Z$ is parity-invariant,  
\beq
\upgamma _{0}Z^{\dagger}\upgamma_{0}=Z,\eeq
it is called a boomerang. Once the Fierz aggregate is constructed, any spinor field $\upzeta\in\mathbb{C}^4$ can be reconstructed from its bilinear covariants up to a phase \cite{Mosna:2003am}. 
This result can be stated by the so-called reconstruction theorem \cite{takas,takas1}. It states that as long as the condition $\upzeta^\dagger\upgamma_0\uppsi\neq 0$ is satisfied, one can express \beq
\uppsi=\frac{1}{2\sqrt{\upzeta{Z}^\dagger\gamma^0\upzeta}}\;e^{-i\upalpha}{Z}\upzeta, \label{31}\eeq where the spinor phase reads $\upalpha=i\ln\left({2}\sqrt{\upzeta^\dagger\gamma^0\uppsi\upzeta^\dagger\gamma^0{Z}\upzeta}\right)$. A particular case states that this result allows for regular spinor fields to be expressed by 
\beq
\uppsi = \frac12\sqrt{S^{12}+J^0+ K^3+\upsigma } {Z} e^{i\upalpha}(1,0,0,0)^\intercal.
 \eeq
 The reconstruction theorem has been generalized for higher dimensions in Ref. \cite{Bonora:2015ppa}.
 
Regular spinor fields satisfy the Fierz--Pauli--Kofink identities, which does not hold, in general, for singular spinor fields. Anyway, the widespread  generalization of the Fierz--Pauli--Kofink identities, holding for both regular and singular spinor fields, takes into account the Fierz aggregate and can be expressed by the following identities,
 \begin{subequations}
\beq
Z\upgamma_{\mu}Z &=&4J_{\mu}Z,\\Z^{2} &=&4\upsigma Z,\\ Zi\upgamma_{\mu}\upgamma_{\nu}Z&=&4S_{\mu\nu}Z,\\
Z\upgamma_0\upgamma_{1}\upgamma_{2}\upgamma_3Z&=&-4\upomega Z,\\ Zi\upgamma_0\upgamma_{1}\upgamma_{2}\upgamma_3\upgamma_{\mu}Z &=&4K_{\mu}Z.
\eeq
\end{subequations}

\section{Classical, Algebraic Spinors and Spinor Operators}
\label{scle} 
Let one considers a multivector $\Omega\in{C}\ell _{1,3}$, expressed as 
\beq
\!\Omega \!=\! c\!+\!\upbeta^{\mu}{\scalebox{0.8}{$\textsc{E}$}}_{\mu}\!+\!\upbeta^{\mu\nu}{\scalebox{0.8}{$\textsc{E}$}}_{\mu}{\scalebox{0.8}{$\textsc{E}$}}_{\nu}\!+\!\upbeta^{\mu\nu\sigma}{\scalebox{0.8}{$\textsc{E}$}}_{\mu}{\scalebox{0.8}{$\textsc{E}$}}_{\nu}{\scalebox{0.8}{$\textsc{E}$}}_{\sigma}\!+\!\upbeta^{0123}{\scalebox{0.8}{$\textsc{E}$}}_{0}{\scalebox{0.8}{$\textsc{E}$}}_{1}{\scalebox{0.8}{$\textsc{E}$}}_{2}{\scalebox{0.8}{$\textsc{E}$}}_{3},
\label{spinor}
\eeq where the anti-symmetry in beta indices are inherited from the base in $Cl_{1,3}$. However, we stress that this situation changes when incorporating nontrivial topology effects, and the resulting coefficients shall no longer be totally anti-symmetric. When taking into account algebraic isomorphism ${C}\ell _{1,3}\simeq {\textsc{M}}(2,\mathbb{H})$, a spinor representation of ${C}\ell _{1,3}$ can be obtained. The idempotent 
$f=\frac{1}{2}(1+{\scalebox{0.8}{$\textsc{E}$}}_{0})$, which is primitive, is an essential ingredient for constructing the left ideal ${C}\ell _{1,3}f$ of ${C}\ell _{1,3}$, which is minimal. A Clifford algebraic element in this ideal reads 
\beq
 \!\!\!\!\!\!\!\!\!\upxi &=&(\upalpha^{1}+\upalpha^{2}{\scalebox{0.8}{$\textsc{E}$}}_{2}{\scalebox{0.8}{$\textsc{E}$}}_{3}+\upalpha^{3}{\scalebox{0.8}{$\textsc{E}$}}_{3}{\scalebox{0.8}{$\textsc{E}$}}_{1}+\upalpha^{4}{\scalebox{0.8}{$\textsc{E}$}}_{1}{\scalebox{0.8}{$\textsc{E}$}}_{2})f\nonumber\\&&+(\upalpha^{5}\!+\!\upalpha^{6}{\scalebox{0.8}{$\textsc{E}$}}_{2}{\scalebox{0.8}{$\textsc{E}$}}_{3}\!+\!\upalpha^{7}{\scalebox{0.8}{$\textsc{E}$}}_{3}{\scalebox{0.8}{$\textsc{E}$}}_{1}\!+\!\upalpha^{8}{\scalebox{0.8}{$\textsc{E}$}}_{1}{\scalebox{0.8}{$\textsc{E}$}}_{2}){\scalebox{0.8}{$\textsc{E}$}}_{0}{\scalebox{0.8}{$\textsc{E}$}}_{1}{\scalebox{0.8}{$\textsc{E}$}}_{2}{\scalebox{0.8}{$\textsc{E}$}}_{3}f.
\label{kkk}
\eeq
\noindent 
Since $\upxi$ and $\Omega f$ are both elements in the minimal left ideal ${C}\ell_{1,3}f$, equalling them yields 
\begin{widetext}
\beq
\upalpha^{1} =c+\upbeta^{0},\qquad \upalpha^{2}=\upbeta^{23}+\upbeta^{023},\qquad \upalpha^{3}=-\upbeta^{13}-\upbeta^{013},\qquad \upalpha^{4}=\upbeta^{12}+\upbeta^{012},\nonumber\\
\upalpha^{5} = -\upbeta^{123}+\upbeta^{0123},\qquad \upalpha^{6}=\upbeta^{1}-\upbeta^{01},\qquad \upalpha^{7}=\upbeta^{2}-\upbeta^{02},\qquad \upalpha^{8}=\upbeta^{3}-\upbeta^{03}.
\eeq\end{widetext}
\noindent 
One can denote the quaternionic units by the bivector fields
\beq
\text{$\imath={\scalebox{0.8}{$\textsc{E}$}}_{2}{\scalebox{0.8}{$\textsc{E}$}}_{3},\qquad\jmath={\scalebox{0.8}{$\textsc{E}$}}_{3}{\scalebox{0.8}{$\textsc{E}$}}_{1},$\qquad $k={\scalebox{0.8}{$\textsc{E}$}}_{1}{\scalebox{0.8}{$\textsc{E}$}}_{2}$}\label{167},\eeq being the set of generators of the quaternion Hurwitz algebra $\mathbb{H}$, together to the unit. The minimal left ideal ${C}\ell_{1,3}f$ is, additionally, a right $\mathbb{H}$-module. One can realize that first expressing the quaternionic coefficients in Eq. (\ref{kkk}) by  
\beq
\mathfrak{q}_1 &=& \upalpha^1+\upalpha^2{\scalebox{0.8}{$\textsc{E}$}}_{2}{\scalebox{0.8}{$\textsc{E}$}}_3+\upalpha^3{\scalebox{0.8}{$\textsc{E}$}}_{3}{\scalebox{0.8}{$\textsc{E}$}}_1+\upalpha^4{\scalebox{0.8}{$\textsc{E}$}}_{1}{\scalebox{0.8}{$\textsc{E}$}}_2, \label{q1}\\\mathfrak{q}_2 &=& \upalpha^5+\upalpha^6{\scalebox{0.8}{$\textsc{E}$}}_{2}{\scalebox{0.8}{$\textsc{E}$}}_3+\upalpha^7{\scalebox{0.8}{$\textsc{E}$}}_{3}{\scalebox{0.8}{$\textsc{E}$}}_1+\upalpha^8{\scalebox{0.8}{$\textsc{E}$}}_{1}{\scalebox{0.8}{$\textsc{E}$}}_2,\label{q2}
\eeq
which lie in the division ring \cite{oxford}
\beq
f{C}\ell_{1,3}f = \textsc{{span}}_\mathbb{R}\{1, {\scalebox{0.8}{$\textsc{E}$}}_{2}{\scalebox{0.8}{$\textsc{E}$}}_3,{\scalebox{0.8}{$\textsc{E}$}}_{3}{\scalebox{0.8}{$\textsc{E}$}}_1,{\scalebox{0.8}{$\textsc{E}$}}_{1}{\scalebox{0.8}{$\textsc{E}$}}_2\} \simeq \mathbb{H}.\eeq Hence the sets $\{\mathfrak{q}_1,\mathfrak{q}_2\}$ and $\{f,{\scalebox{0.8}{$\textsc{E}$}}_{0}{\scalebox{0.8}{$\textsc{E}$}}_1{\scalebox{0.8}{$\textsc{E}$}}_{2}{\scalebox{0.8}{$\textsc{E}$}}_3\}$ do commute. Therefore one concludes that the minimal left ideal ${{C}\ell_{1,3}}f$ consists of a (right) $\mathbb{H}$-module over the division ring $\mathbb{H}$, having a complete basis given by the set $\{f, {\scalebox{0.8}{$\textsc{E}$}}_{0}{\scalebox{0.8}{$\textsc{E}$}}_{1}{\scalebox{0.8}{$\textsc{E}$}}_{2}{\scalebox{0.8}{$\textsc{E}$}}_{3}f\}$.

The following matrix representations can be straightforwardly obtained, 
\begin{equation}
{\scalebox{0.8}{$\textsc{E}$}}_{0}=%
{\small\begin{bmatrix}
1 & 0 \\ 
0 & -1%
\end{bmatrix}}%
,\quad{\scalebox{0.8}{$\textsc{E}$}}_{1}=%
{\small\begin{bmatrix}
0 & \imath \\ 
\imath & 0%
\end{bmatrix}}%
,\quad{\scalebox{0.8}{$\textsc{E}$}}_{2}=%
{\small\begin{bmatrix}
0 & \jmath \\ 
\jmath & 0%
\end{bmatrix}}%
,\quad{\scalebox{0.8}{$\textsc{E}$}}_{3}=%
{\small\begin{bmatrix}
0 & k \\ 
k & 0%
\end{bmatrix}}\,.%
\end{equation}%
\noindent 
Consequently, the representations of the primitive idempotent $f$ and the volume element ${\scalebox{0.8}{$\textsc{E}$}}_{0}{\scalebox{0.8}{$\textsc{E}$}}_1{\scalebox{0.8}{$\textsc{E}$}}_2{\scalebox{0.8}{$\textsc{E}$}}_3f$ respectively read,  
$$
[f]={\small\begin{bmatrix}
1 & 0 \\ 
0 & 0%
\end{bmatrix}}, \quad \quad 
[{\scalebox{0.8}{$\textsc{E}$}}_{0}{\scalebox{0.8}{$\textsc{E}$}}_{1}{\scalebox{0.8}{$\textsc{E}$}}_{2}{\scalebox{0.8}{$\textsc{E}$}}_{3}f]=%
{\small\begin{bmatrix}
0 & 0 \\ 
1 & 0%
\end{bmatrix}}.
$$ 
In this way, a Clifford multivector representation $\Omega\in{C}\ell_{1,3}$ reads 
\begin{widetext}
\begin{eqnarray} \label{quat}
\!\!\!\!\!\!\!\!\!\!\!\!\!\!\!\Omega &=& \left[%
\begin{array}{cc}
\begin{array}{c}
c + \upbeta^0 + (\upbeta^{23} + \upbeta^{023})\imath \\ 
-(\upbeta^{13} + \upbeta^{013})\jmath + (\upbeta^{12} + \upbeta^{012})k \\ 
\quad\quad\quad\quad\quad\quad\quad \\ 
(\upbeta^{0123}-\upbeta^{123}) + (\upbeta^1 - \upbeta^{01})\imath \\ 
+(\upbeta^2 - \upbeta^{02})\jmath + (\upbeta^3 - \upbeta^{03})k%
\end{array}
\begin{array}{c}
-(\upbeta^{123} + \upbeta^{0123}) + (\upbeta^1 + \upbeta^{01})\imath + \\ 
(\upbeta^2 + \upbeta^{02})\jmath + (\upbeta^3 + \upbeta^{03})k \\ 
\quad\quad\quad\quad\quad\quad\quad \\ 
(c - \upbeta^0) + (\upbeta^{23} - \upbeta^{023})\imath + \\ 
(\upbeta^{013}-\upbeta^{13})\jmath + (\upbeta^{12} - \upbeta^{012})k%
\end{array}
& 
\end{array}%
\right] = {\small\begin{bmatrix}
\mathfrak{q}_1 & \mathfrak{q}_2 \\ 
\mathfrak{q}_3 & \mathfrak{q}_4%
\end{bmatrix}}.
\end{eqnarray}
\end{widetext}
When one considers the even Clifford subalgebra, a {spinor operator} $\uppsi \in {C}\ell_{1,3}^{+}$ has matrix representation given by 
\begin{equation}
[\uppsi]=c+\upbeta^{\mu\nu}{\scalebox{0.8}{$\textsc{E}$}}_{\mu}{\scalebox{0.8}{$\textsc{E}$}}_{\nu}+\upbeta^{0123}{\scalebox{0.8}{$\textsc{E}$}}_{0}{\scalebox{0.8}{$\textsc{E}$}}_{1}{\scalebox{0.8}{$\textsc{E}$}}_{2}{\scalebox{0.8}{$\textsc{E}$}}_{3}\,,
\label{400}
\end{equation}%
\noindent which can be equivalently expressed as
\begin{widetext} 
\begin{equation}
{\small\begin{bmatrix}
\mathfrak{q}_{1} & -\mathfrak{q}_{2} \\ 
\mathfrak{q}_{2} & \mathfrak{q}_{1}%
\end{bmatrix}}%
= 
\begin{bmatrix}
c+\upbeta^{23}\imath-\upbeta^{13}\jmath+\upbeta^{12}k & \quad
-\upbeta^{0123}+\upbeta^{01}\imath+\upbeta^{02}\jmath+\upbeta^{03}k \\ 
\upbeta^{0123}-\upbeta^{01}\imath-\upbeta^{02}\jmath-\upbeta^{03}k & \quad
c+\upbeta^{23}\imath-\upbeta^{13}\jmath+\upbeta^{12}k
\end{bmatrix} . 
\end{equation}
\end{widetext}
The isomorphisms\footnote{The last two isomorphisms are isomorphisms only among vector spaces.} ${C}\ell _{1,3}^{+} \simeq {C}\ell _{1,3}\frac{1}{2}(1+{\scalebox{0.8}{$\textsc{E}$}}_{0})\simeq \mathbb{C}^{4}\simeq \mathbb{H}^{2}$, therefore, illustrate the
three equivalent definitions of a spinor field, namely, the algebraic spinor, the spinor operator, and the classical one. Hence, the classical spinor space of Minkowski spacetime, $\mathbb{H}^{2}$, carrying the  
${\scalebox{0.8}{D}}^{(1/2,0)}\oplus {\scalebox{0.8}{D}}^{(0,1/2)}$ of the Lorentz group, is thus isomorphic to the (minimal left) ideal 
${C}\ell _{1,3}\frac{1}{2}(1+{\scalebox{0.8}{$\textsc{E}$}}_{0})$, whose corresponding representation is carried by the algebraic spinor. This ideal is also isomorphic to the even Clifford subalgebra ${C}\ell _{1,3}^{+}$, whose associated ideal ${C}\ell_{1,3}f\simeq \mathbb{H}\oplus \mathbb{H}$ has arbitrary elements given by
\begin{equation}
{\small\begin{bmatrix}
\mathfrak{q}_{1} & -\mathfrak{q}_{2} \\ 
\mathfrak{q}_{2} & \mathfrak{q}_{1}%
\end{bmatrix}} [f]
=
\begin{bmatrix} \mathfrak{q}_1 & -\mathfrak{q}_2\\\mathfrak{q}_2 & \phantom{-}\mathfrak{q}_1 \end{bmatrix}
 \begin{bmatrix} 1 & 0\\0 & 0 \end{bmatrix}
\!=\! {\small\begin{bmatrix} \mathfrak{q}_1 & 0\\\mathfrak{q}_2 & 0 \end{bmatrix}}\simeq{\small\begin{bmatrix} \mathfrak{q}_1\\\mathfrak{q}_2 \end{bmatrix}} 
\!=\!\left[ 
\begin{array}{cc}
c\!+\!\upbeta^{23}\imath\!-\!\upbeta^{13}\jmath\!+\!\upbeta^{12}k&0 \\ 
\upbeta^{0123}\!-\!\upbeta^{01}\imath\!-\!\upbeta^{02}\jmath-\upbeta^{03}k&0%
\end{array}\right].\label{hh}
\end{equation}%
\noindent 
Returning to Eqs.~(\ref{167}, \ref{400}), when the representation 
\begin{gather}
1 \mapsto {\small\begin{bmatrix} 1 & 0\\ 0 & 1\end{bmatrix}}, \quad
\ii \mapsto {\small\begin{bmatrix} i & \phantom{-}0\\ 0 & -i \end{bmatrix}}, \quad
\jj \mapsto {\small\begin{bmatrix} \phantom{-}0 & 1\\ -1 & 0 \end{bmatrix}}, \quad
\kk \mapsto {\small\begin{bmatrix} 0 & i\\ i & 0 \end{bmatrix}}\,, 
\label{quatrepr}
\end{gather}
is taken into account, for $i$ denoting the imaginary unit, the spinor operator $\uppsi$ representation in Eq. (\ref{400}) reads  
\begin{widetext}
\begin{equation}
\begin{bmatrix} 
c+\upbeta^{23}i & -\upbeta^{13}+\upbeta^{12}i & -\upbeta^{0123}+\upbeta^{01}i & \upbeta^{02}+\upbeta^{03}i \\
\upbeta^{13}+\upbeta^{12}i & c-\upbeta^{23}i & -\upbeta^{02}+\upbeta^{03}i & -\upbeta^{0123}-\upbeta^{01}i\\
\upbeta^{0123}-\upbeta^{01}i & -\upbeta^{02}-\upbeta^{03}i & c+\upbeta^{23}i & -\upbeta^{13}+\upbeta^{12}i\\
\upbeta^{02}-\upbeta^{03}i & \upbeta^{0123}+\upbeta^{01}i & \upbeta^{13}+\upbeta^{12}i & c-\upbeta^{23}i\\
\end{bmatrix} 
\equiv%
\begin{bmatrix} 
\upphi_1 & -\upphi_2^{*} & -\upphi_3 & \phantom{-}\upphi_4^{*}\\
\upphi_2 & \phantom{-}\upphi_1^{*} & -\upphi_4 & -\upphi_3^{*}\\
\upphi_3 & -\upphi_4^{*} & \phantom{-}\upphi_1 & -\upphi_2^{*}\\
\upphi_4 & \phantom{-}\upphi_3^{*} & \phantom{-}\upphi_2 & \phantom{-}\upphi_1^{*}
\end{bmatrix}.
\end{equation}
\end{widetext}
The usual Dirac spinor $\uppsi $ resides in the complexified ideal $(\mathbb{C}\otimes \cl _{1,3})f$, generated by the primitive idempotent $ f=\frac{1}{4}(1+{\scalebox{0.8}{$\textsc{E}$}}_{0})(1+i{\scalebox{0.8}{$\textsc{E}$}}_{1}{\scalebox{0.8}{$\textsc{E}$}}_{2})$, which originates the Dirac representation of gamma matrices as endomorphisms of $\mathbb{C}^4$ \cite{oxford}. One can choose still the primitive idempotent $f=\frac{1}{4}(1+i{\scalebox{0.8}{$\textsc{E}$}}_{0}{\scalebox{0.8}{$\textsc{E}$}}_{1}{\scalebox{0.8}{$\textsc{E}$}}_{2}{\scalebox{0.8}{$\textsc{E}$}}_{3})(1+i{\scalebox{0.8}{$\textsc{E}$}}_{1}{\scalebox{0.8}{$\textsc{E}$}}_{2})$ to engender the Weyl representation, whereas the Majorana representation is constructed uppon the choice of the primitive idempotent $f=\frac{1}{4}(1+i{\scalebox{0.8}{$\textsc{E}$}}_{1})(1+i{\scalebox{0.8}{$\textsc{E}$}}_{0}{\scalebox{0.8}{$\textsc{E}$}}_{2})$. 
Each one of those possible representations is denoted by 
$\upgamma(\scalebox{0.8}{$\textsc{E}$}_\mu):=\upgamma_{\mu}$, 
where $\upgamma: \mathbb{C}\otimes \cl _{1,3}\to \mathrm{End}(\mathbb{C}^{4})$ is the Clifford representation into the endomorphism space of $\mathbb{C}^{4}$. 

In this way, the algebraic Dirac spinor field can be therefore expressed, in the Dirac representation of gamma matrices, by \beq
\uppsi =\Omega \frac{1}{2}(1+i\upgamma _{1}\upgamma_2)\in (\mathbb{C}\otimes \cl _{1,3})f,\eeq where 
\beq\Omega =[\uppsi] \frac{1}{2}(1+\upgamma _{0})\in {C}\ell_{1,3}(1+\upgamma _{0}) = 2\Re(\uppsi).\eeq Hence 
the algebraic spinor field representation  
\beq
\uppsi \simeq  
\begin{bmatrix}\label{rep1}
\upphi_1 & 0 & 0 & 0 \\ 
\upphi_2 & 0 & 0 & 0 \\ 
\upphi_3 & 0 & 0 & 0 \\ 
\upphi_4 & 0 & 0 & 0%
\end{bmatrix}\in (\CC\otimes \cl _{1,3})f,
\eeq
is isomorphic to the usual 4-spinor 
\beq
\begin{bmatrix}
\upphi_1 \\ 
\upphi_2 \\ 
\upphi_3 \\ 
\upphi_4%
\end{bmatrix}%
=%
\begin{bmatrix}
\uppsi_1 \\ 
\uppsi_2 \\ 
\uppsi_3 \\ 
\uppsi_4%
\end{bmatrix}%
\in \mathbb{C}^{4} \label{dire}
\eeq
and its interpretation as a classical spinor field definition can be immediately read off.

\section{Merging the Cartan's spinor formalism and exotic spinors}
\label{nv}

In the presence of a base manifold endowed with nontrivial topology, more than one spin structure is allowed. These different spin structures comprise different spinors, the so-called exotic spinors. Therefore, if we call $\psi$ a section of $\mathbb{P}_{\mathrm{Spin}_{1,3}^{e}}(M)\times_\rho \mathbb{C}^4$, as before, then there exists another vector bundle $\overcirc{\mathbb{P}}_{\mathrm{Spin}_{1,3}^{e}}(M)\times_\rho \mathbb{C}^4$ of which exotic spinors, $\overcirc{\psi}$ are sections. Several works well describe the relationship between both bundles, which will not be repeated here \cite{petry,field,f2,exotic}. It suffices for our purposes to describe how derivative operators act upon exotic spinors. For that, we start with the fact that there is a (matrix, linear and invertible) mapping, say $\kappa$, between exotic and usual spinors:
\begin{eqnarray}
\kappa:\overcirc{\mathbb{P}}_{\mathrm{Spin}_{1,3}^{e}}(M)\times_\rho &&\left.\mathbb{C}^4 \rightarrow \mathbb{P}_{\mathrm{Spin}^e_{1,3}}\times_\rho \mathbb{C}^4\nonumber\right.\\&&\left.\overcirc{\psi}\mapsto \kappa\overcirc{\psi}=\psi.\right.\label{mmap}
\end{eqnarray} We can work out the correction in the derivative of $\overcirc{\psi}$ due to the mapping $\kappa$, very much like the appearance of gauge potentials and covariant derivatives. To save notation, let $d$ be a spinorial derivative operator. Eq. (\ref{mmap}) yields $d\psi=d\kappa \overcirc{\psi}+\kappa d \overcirc{\psi}$, and the insertion of $\kappa^{-1}$ from the left leads to 
\begin{equation}
d\overcirc{\psi}=\kappa^{-1} d\psi-\kappa^{-1}d\kappa \overcirc{\psi}. \label{dpsi}
\end{equation} Let one calls $\mathcal{A}$ the derivative operator correction when acting upon an exotic spinor so that $\overcirc{d}\overcirc{\psi}=(d+\mathcal{A})\overcirc{\psi}$. Using (\ref{dpsi}) and inserting $\kappa$ from the left implies that  
\begin{equation}
\kappa \overcirc{d}\overcirc{\psi}=d\psi+\kappa(\mathcal{A}-\kappa^{-1}d\kappa)\overcirc{\psi}.\label{qqq}
\end{equation} The derivative of a spinor is also a spinor; hence in order to have $\kappa \overcirc{d}\overcirc{\psi}=d\psi$ (just as for $\kappa\overcirc{\psi}=\psi$) we recognize $\mathcal{A}=\kappa^{-1}d\kappa$, leading to $\overcirc{d}=d+\kappa^{-1}d\kappa$ as the exotic derivative operator. Apart from these similarities with a gauge potential, there is a crucial difference: $\kappa$ are functions whose roots express the m\"obiusity of the nontrivial base manifold, and, as such, they present discontinuity at given surfaces. There is no globally defined gauge potential, obviously, with such behavior. We finalize these comments emphasizing that the derivative operator corrections act, ultimately, upon spinor entries and, therefore, it is important to review some relevant points about it in Cartan's spinor formalism.   

As a standard procedure to Cartan's approach, unrelated to nontrivial topology, the construction starts by slicing the light-cone at a given fixed time. One then arrives at the so-called celestial sphere, $S$ \cite{cart,pen}. Let $S^+$ be the upper half of $S$ when intercepted by the complex plane at the equator. Stereographic projections of $S^+$ onto the complex plane present the inconvenience of an absence of a regular range for the north pole. In fact, calling $\varphi$ the stereographic projection $\varphi:S^+\rightarrow \mathbb{C}^2$ and $N$ the north pole in $S^+$, it can be readily verified that $\varphi(N)=\infty$. This problem is addressed with the aid of two variables over the complex field such that $\varphi(\cdot)=\zeta/\xi|_{(.)}$. When denoted as $(\zeta,\,\xi)$, these complex coordinates engendering the stereographic projection encompass the north pole simply by $(1,\, 0)$. As $\zeta$ and $\xi$ map points of $S^+$, they may be written in terms of spacetime coordinates. In this vein, denoting by $M=(\zeta,\, \xi)^\intercal(\zeta^*,\, \xi^*)$ one has $\det(M)=t^2-x^2-y^2-z^2$, where the speed of light $c$ has been set equal to unit. 

The pair $\zeta$ and $\xi$ are the spinor entries within Cartan's formalism and, more importantly to our purposes, a given spacetime point $P=(t,x,y,z)$ may be expressed as $\sqrt{2}P=(\zeta\zeta^*+\xi\xi^*, \zeta\xi^*+\xi\zeta^*,i(\xi\zeta^*-\zeta\xi^*), \zeta\zeta^*-\xi\xi^*)$, where $\alpha^*$ means complex conjugate of $\alpha$. In the sense of this last remark, one usually says that spinors are the square root of geometry. We now turn to motivate the correction of the dual coefficients by taking both formalisms altogether. All the details are discussed in Ref. \cite{JHEP}, and here only the appropriate steps for our purposes will be provided.   

Let us start with the usual procedure to get the differentials over $\mathbb{R}^n$. Let $\{e_i\}$ be a basis of $\mathbb{R}^n$ and $\{dx^j\}$ a basis of $(\mathbb{R}^n)^*$, so that $dx^j(e_i)=\delta^j_{\;i}$. Besides, let $h$ be a vector field in $\mathbb{R}^n$, and $x_0$ be a point within a given open set of $\mathbb{R}^n$. The differential of a given funcion $f$ in $x_0$ acting upon $h$ reads $df(x_0)(h)=\sum_i[\partial f/\partial x^i]_{x_0}h^i$. To bring now this expression to a usual form, one takes the linear orthogonal projections $\pi^i:\mathbb{R}^n\rightarrow \mathbb{R}$, i.e., $\pi^i(x^1,\ldots,x^n)=x^i$ so that $d\pi(h)=dx^i(h)=h^i$ and hence $df(x_{0})=\sum\limits_{i}\frac{\partial f}{\partial x^{i}}(x_{0})dx^{i}$ for all $h$. Nontrivial topology effects may be introduced through a peculiar distinction between $d\pi^i$ and $dx^i$ taking into account Cartan's formalism and exotic spinors, for, since now $x^\mu\sim (\xi\xi^*)^\mu$, as mentioned in the previous paragraph, yields  
\begin{equation}\label{indo}
 d\pi^{\mu}\!=\!d(\zeta^*\zeta)^{\mu}\!=\!\sum_\nu\frac{\partial}{\partial x_\nu}(\zeta^*\zeta)^{\mu}dx^\nu\!=\!\sum_\nu(\zeta^*\partial_\nu\zeta+\partial_\nu\zeta^*\zeta)^{\mu}dx^\nu. 
 \end{equation} Motivated by the discussion around Eq. (\ref{qqq}), the derivatives acting upon the spinor entries $\xi$ within the nontrivial topology context will be shifted with a rescaling $\partial_\mu\mapsto \partial_\mu+\partial_\mu\theta$. With this, a simple rescaling in the $\theta(x)$ function leads to\footnote{As in Ref. \cite{JHEP}, we shall deal with $\theta\in\mathbb{R}$ here. The full exploration of all possibilities for this shifting term shall be explored in a forthcoming publication.}  
\begin{eqnarray}
d\pi^{\mu}=dx^{\mu}+x^{\mu}d\theta
\end{eqnarray} and differentials now should read
 \begin{equation}\label{df corrigido}
df(x_{0})
=
\sum\limits_{\mu}
\frac{\partial f}{\partial x^{\mu}}(x_{0})dx^{\mu}
+
\sum\limits_{\mu}\frac{\partial f}{\partial x^{\mu}}(x_{0})x^{\mu}d\theta.
\end{equation} 
 
Given this, a corrected quadratic form, carrying signatures of the nontrivial topology, reads 
\begin{eqnarray}\label{aq}
g=\eta_{\mu\nu}(dx^\mu+x^\mu d\theta)\otimes(dx^\nu+x^\nu d\theta)\equiv g_{\mu\nu}{\scalebox{0.8}{$\textsc{E}$}}^{\mu }\otimes{\scalebox{0.8}{$\textsc{E}$}}^{\nu }, 
\end{eqnarray} whose action upon base vectors leads to 
\beq\label{eq2}
g_{\mu\nu}=\eta_{\mu\nu}+x_{(\mu}\partial_{\nu)}\theta+x^2\partial_\mu\theta\partial_\nu\theta,\eeq where $x_{(\mu}\partial_{\nu)}\theta\equiv x_\mu\partial_\nu\theta+x_\nu\partial_\mu\theta$. Now it is appropriate an attempt to interpret the $\theta$-corrections. There are two usual ways to figure out such terms appearing in the metric. One could think of such a correction in order when the topology is locally nontrivial in a given limited spacetime region. Equivalently, notice that in natural units (with $\theta$ dimensionless) scales as $energy$ or $(lenght)^{-1}$. Therefore, the $\theta$-correction terms would enter as a high-energy effect onto spacetime. In any case, we will work here in a setup in which first derivative products, powers of $\theta$, and higher order $\theta$ derivatives are to be disregarded. Although many of the results hereon presented are independent of these approximations, it helps to be supported by them to make explicit the equivalence of the methods here used with the ones of Ref. \cite{JHEP}.  

\section{Extended Algebraic Spinor Fields}
\label{scle1}
Let one denotes by $\cl(M,\eta)$ the Clifford bundle over the four-dimensional manifold $M$, endowed with a symmetric bilinear form $\eta: \Upomega(M)\times \Upomega(M)\rightarrow\mathbb{R}$, up to now considered as the Clifford bundle of the Minkowski spacetime \cite{Mosna:2002fr}. The metric $\eta$ defines a quadratic form $Q: \Upomega^{1}(M)\rightarrow\mathbb{R}$. One also denotes by $\cl(M,g)$ the Clifford bundle constructed over~$M$ equipped with the exotic bilinear form presented in the previous section, $
g = g_{\mu\nu}{\scalebox{0.8}{$\textsc{E}$}}^{\mu }\otimes{\scalebox{0.8}{$\textsc{E}$}}^{\nu }$,
where
\beq
\label{exo}
g_{\mu\nu}=\eta_{\mu\nu}+x_{(\mu}\partial_{\nu)}\theta.\eeq 
It is worth emphasizing that the manifold $M$, being multiply connected, has a non-trivial topology. Since it presents more than one spin structure, the $\cl(M,\eta)$ Clifford bundle can be taken as the one related to the trivial spin structure, whereas the Clifford bundle $\cl(M,g)$ carries the nontrivial topological effects through Eq. (\ref{exo}). 
In fact, since $Q(u) = \eta(u, u)$, for $u\in \sec\Upomega^{1*}(M)$, the bundle $\cl(M,\eta)$ is isomorphic to the bundle $\cl(M,g)$, whose underlying vector space is $\Upomega(M)$, with the (associative) Clifford product defined as\footnote{Hereon the notation $\uppsi$ will be used for denoting a differential form. Although this notation has been used for denoting a spinor field in the previous section, the use of the same notation will be clear, after Eq. (\ref{spinorq}).}
\beq
u\pu \uppsi = u \wedge \uppsi + u\JJBB \uppsi\label{b11},\eeq
where $\uppsi \in \sec\Upomega(M)$ and $\displaystyle{u\JJBB \uppsi}$ is the left contraction which is an (anti-)derivation of degree $-1$ defined on $\Upomega(M)$ and defined on the cotangent space as \beq
u\displaystyle{\JJBB} v = g(u, v)\mathbb{I}_{4},\eeq where $v \in \sec\Upomega^{1*}(M)$ is an arbitrary vector field.
Hence, the $g$-product (\ref{b11}), in the Clifford bundle context, makes $\cl(M,\eta)$ and $\cl(M,g)$ to be isomorphic. Besides, the Clifford bundle $\cl(M,g)$ is led to $\cl(M,\eta)$ on Minkowski space when $\theta$ is constant in Eq. (\ref{exo}). Although 
$\cl(M,g)$ and $\cl(M,\eta)$ are isomorphic, physical signatures of the nontrivial topology can be derived and analyzed when 
the exotic Clifford bundle $\cl(M,g)$ is looked at from the point of view of $\cl(M,\eta)$, besides some dualities, the inner representation structure, and mainly the concept of vacuum structure as the unique algebraic projection on the base field immersed in the Clifford bundle, being equivalent to the vacuum in QFT and the Gelfand--Naimark--Segal \cite{Ablamowicz:2014rpa}, are different for the two Clifford bundles $\cl(M,g)$ and $\cl(M,\eta)$.

The $g$-bilinear covariants for the basis  
\beq
\!\!\!\mathcal{B} \!=\! \{1, {\scalebox{0.8}{$\textsc{E}$}}_\mu, {\scalebox{0.8}{$\textsc{E}$}}_\mu\wedge {\scalebox{0.8}{$\textsc{E}$}}_\nu, {\scalebox{0.8}{$\textsc{E}$}}_\mu\wedge {\scalebox{0.8}{$\textsc{E}$}}_\nu\wedge {\scalebox{0.8}{$\textsc{E}$}}_\sigma, {\scalebox{0.8}{$\textsc{E}$}}_{0}\wedge {\scalebox{0.8}{$\textsc{E}$}}_1 \wedge {\scalebox{0.8}{$\textsc{E}$}}_2 \wedge {\scalebox{0.8}{$\textsc{E}$}}_3\}
\eeq
should be taken into account to performing measurements. 

Intending to introduce bilinear covariants for extended Clifford bundles, one again considers the tangent space at a point $x\in M$, namely, $\Upomega^{1*}(T_xM)\simeq\mathbb{R}^{1,3}$. The $g$-products between homogeneous multivector fields and arbitrary multivector fields are defined as follows:
 \begin{enumerate}
\item When a vector $u\in V$ is taken into account, for any $\uppsi\in \sec \cl(M,g)$, one can denote  
\beq\label{01}u\pu\uppsi &=& u\wedge\uppsi + u\JJgg\uppsi = 
u\wedge\uppsi + (u\uppsi)_\theta + u\JJee\uppsi\nonumber\\&=& u\uppsi + (u\uppsi)_\theta.
\eeq
For the particular case where $u = \upgamma_\mu$ and $\uppsi = \upgamma_\nu$, one has
\beq
\upgamma_\mu\JJgg\upgamma_\nu = g(\upgamma_\mu,\upgamma_\nu)=g_{\mu\nu}=\eta_{\mu\nu}+x_{(\mu}\partial_{\nu)}\theta,
\eeq
or, equivalently,  
\beq\label{38}
\upgamma_\mu\JJgg\upgamma_\nu = \upgamma_\mu\JJee\upgamma_\nu + x_{(\mu}\partial_{\nu)}\theta.
\eeq
In this regard, we call attention to the implementation of nontrivial topology correction via left contraction, contrasting it with the approach presented in \cite{JHEP}. Recall that the Clifford product acting upon a given basis reads
\beq
\gamma(e_\mu)\gamma(e_\nu)=\gamma(e_\mu)\wedge\gamma(e_\nu)+\gamma(e_\mu)\JJgg\gamma(e_\nu),
\eeq so that, in view of Eq. (\ref{38}), one has for the anticommutator $\{\gamma(e_\mu),\gamma(e_\nu)\}=2\big(\eta_{\mu\nu}+ x_{(\mu}\partial_{\nu)}\theta\big)\mathbb{I}_{4}=2g_{\mu\nu}\mathbb{I}_{4}$, as expected. In Ref. \cite{JHEP}, the gamma matrices of the corrected spacetime were related to the usual Minkowskian ones, $\Gamma$, by $\gamma(e^\mu)=(\delta^\mu_\alpha-x^\mu\partial_\alpha\theta)\Gamma(e^\alpha)$, where naturally $\{\Gamma(e^\alpha), \Gamma(e^\beta)\} =2\eta^{\alpha\beta}$. Using the approximations discussed around Eq. (\ref{aq}), it is not difficult to see that  
\beq
\gamma(e^\mu)\gamma(e^\nu)&=&g^{\mu\nu}+\Gamma(e^\mu)\wedge\Gamma(e^\nu)-\partial_\alpha\theta\Gamma(e^\alpha)\wedge(x^\mu\Gamma(e^\nu)\!-\!x^\nu\Gamma(e^\mu))
\eeq from which the Clifford algebra relation is straightforwardly recovered. However, the advantage of implementing corrections via left contraction manifests in the cases related below. The notation  
\beq
\Uptheta_{\mu\nu}=x_{(\mu}\partial_{\nu)}\theta
\eeq
will be used hereon. 
\item 
One can consider the Clifford product between a 2-vector $uv$ and a multivector $\uppsi$, whose particular case where $u = \upgamma_\mu$, $v = \upgamma_\nu$, and $\uppsi = \upgamma_\rho$ yields 
\beq
{\upgamma_\mu}\pu{\upgamma_\nu} \pu {\upgamma_\rho} &=& \upgamma_\mu\wedge\upgamma_\nu \wedge \upgamma_\rho
+ g_{\mu\nu}\upgamma_\rho - g_{\mu\rho}\upgamma_\nu+g_{\nu\rho}\gamma_\mu\nonumber\\
&=& \upgamma_\mu\wedge\upgamma_\nu \wedge \upgamma_\rho
+ (\eta_{\mu\nu}+\Theta_{\mu\nu})\upgamma_\rho - (\eta_{\mu\rho}+\Theta_{\mu\rho})\upgamma_\nu+(\eta_{\nu\rho}+\Theta_{\nu\rho})\gamma_\mu\nonumber\\
&=&  \upgamma_\mu\upgamma_\nu \upgamma_\rho +\Theta_{\mu\nu}\upgamma_\rho - \Theta_{\mu\rho}\upgamma_\nu +  \Theta_{\nu\rho}\gamma_\mu.\label{cfl}
\eeq
\noindent The first term in Eq. (\ref{cfl}) denotes the standard Clifford product in Minkowski spacetime
\beq
 \upgamma_\mu\upgamma_\nu \upgamma_\rho = \upgamma_\mu\wedge\upgamma_\nu \wedge \upgamma_\rho
+ \eta_{\mu\nu}\upgamma_\rho - \eta_{\mu\rho}\upgamma_\nu+\eta_{\nu\rho}\gamma_\mu.\label{etad}
 \eeq
 \end{enumerate}
Eqs. (\ref{01}) -- (\ref{cfl}) are sufficient to define what we have called `extended' spinor fields in exotic Clifford bundles. A word is necessary about this concept: due to the nontrivial topology, the underlying spacetime is not purely the Minkowski one. Therefore its symmetry group is not the Poincar\'e anymore. The spinor here is a deformed object that falls into the usual case (a mathematical quantity carrying the spin-$1/2$ representation of the Spin group) when the topology is trivial. That is what is called here an extended $g$-spinor field. Of course, related terms (such as extended Clifford bundle) are readily accomplished.  

Eq.~(\ref{dire}) utilized the minimal left ideal generated by the primitive idempotent, 
\beq
f=\frac{1}{4}(1+\upgamma_0)(1+i\upgamma_1\upgamma_2)
 = \frac14(1+\upgamma_0 +i\upgamma_1\upgamma_2 +i\upgamma_0\upgamma_1\upgamma_2).\label{ug0}
 \eeq
 From the exotic Clifford bundle $\cl(M,g)$ point of view, the formalism developed in the previous section can be straightforwardly recovered, when now the left minimal ideal $\cl(M,g)f_g$ is regarded, where 
\beq
\label{fb} 
f_g=\frac{1}{4}(1+\upgamma_0)\pu(1+i\upgamma_1\pu{\upgamma_2})\label{012}\,.
\eeq
Therefore, the development paved for $\cl(M,\eta)$ can be borrowed, \emph{mutatis mutandis}, when one modifies the usual Clifford product
$\upgamma_\mu\upgamma_\nu$, introducing the extended Clifford product, 
\beq
\label{q1}
\upgamma_\mu\upgamma_\nu\mapsto\upgamma_\mu\underset{g}{}\upgamma_\nu=\upgamma_\mu\upgamma_\nu + x_{(\mu}\partial_{\nu)}\theta.
\eeq

The last expression is the prominent essence of transliterating $\cl(M, g)$ to $\cl(M,\eta)$. In the extended Clifford bundle $\cl(M,g)$, the left minimal ideal $\cl(M,g)f_g$, that is generated by the extended primitive idempotent (\ref{fb}), can be expressed by 
\begin{eqnarray}
f_g &\!=\!& \frac{1}{4}(1\!+\!\upgamma_0)(1\!+\!i\upgamma_1\upgamma_2) +f(\Uptheta)\,,
\label{f11}
\end{eqnarray}
where \beq
{f}(\Uptheta) = \frac{i}{4}\left[x_{(1}\partial_{2)}\theta(1+\upgamma_0) \!-\! x_{(2}\partial_{0)}\theta\upgamma_1 \!+\! x_{(0}\partial_{1)}\theta\upgamma_2\right].\eeq This approach is independent of the representation chosen. 
Substituting the expression  
\beq
\upgamma_0\underset{g}{}\upgamma_1\underset{g}{}\upgamma_2 = \upgamma_0\upgamma_1\upgamma_2 + \Uptheta_{01}\upgamma_2 - \Uptheta_{02}\upgamma_1 + \Uptheta_{12}\upgamma_0,\label{0122}
\eeq
 into~(\ref{f11}) yields (in the Dirac representation) 
\begin{widetext}
\beq
f_g =  
\overbrace{\begin{bmatrix}1&0&0&0\\0&0&0&0\\
0&0&0&0\\0&0&0&0\end{bmatrix}}^{f} +
\overbrace{\frac{1}{4}\begin{bmatrix}2i\Uptheta_{12}&0&0&-i\Uptheta_{20}-\Uptheta_{01}\\0&2i\Uptheta_{12}&-i\Uptheta_{20}+\Uptheta_{01}&0\\
0&i\Uptheta_{20}+\Uptheta_{01}&0&0\\i\Uptheta_{20}-\Uptheta_{01}&0&0&0
\end{bmatrix}}^{{f}(\Uptheta)}.
\eeq
\end{widetext}
 The case $\Uptheta_{\mu\nu} = 0$ implies $g= \eta$, and the standard spinor formalism can be recovered. 

Now one considers an arbitrary extended multivector field $\uppsi_g\in\sec \cl(M,g)=\sec\clb$, 
\begin{eqnarray}
\uppsi_g \!=\! c\!+\!\upbeta^{\mu}\upgamma_{\mu}\!+\!\upbeta^{\mu\nu}\upgamma_{\mu}\pu{\upgamma_\nu}\!+\!\upbeta^{\mu\nu\sigma}\upgamma_{\mu}\pu{\upgamma_\nu}\pu{\upgamma_{\sigma}}\!+\!\upbeta^{0123}\upgamma_{0}\pu{\upgamma_{1}}\pu{\upgamma_{2}}\pu{\upgamma_{3}}. 
\label{spinorq}
\end{eqnarray}%
Employing Eqs.~(\ref{012},\ref{q1}) and denoting the pseudoscalar coefficient by $p=\upbeta^{0123}$, one can express the last expression as 
\begin{eqnarray}
\uppsi_g &=& c+\upbeta^\mu\upgamma_\mu + \upbeta^{\mu\nu}\upgamma_{\mu}\pu{\upgamma_{\nu}} + \upbeta^{\mu\nu\rho}\upgamma_{\mu}\pu{\upgamma_{\nu}}\pu{\upgamma_\rho} + p\upgamma_0\pu{\upgamma_1}\pu{\upgamma_2}\pu{\upgamma_3}\nonumber\\
 &\!=\!& \uppsi + \upbeta^{\mu\nu}\Uptheta_{\mu\nu} \!+\! \upbeta^{\mu\nu\rho}(\Uptheta_{\mu\nu}\upgamma_\rho - \Uptheta_{\rho\mu}\upgamma_\nu \!+\! \Uptheta_{\nu\rho}\upgamma_\mu)
\nonumber\\&+&p(\Uptheta_{23}\upgamma_{0}\upgamma_{1} \!-\! \Uptheta_{13}\upgamma_{0}\upgamma_{2} \!+\! \Uptheta_{12}\upgamma_{0}\upgamma_{3} \!+\! \Uptheta_{03}\upgamma_{1}\upgamma_{2} \!-\! \Uptheta_{02}\upgamma_{1}\upgamma_{3} \!+\! \Uptheta_{01}\upgamma_{2}\upgamma_{3},  \label{psib}
\end{eqnarray}
and $\uppsi$ is an arbitrary multivector in $\cle$, as presented in Eq. (\ref{spinor}). The notation in Eq.~(\ref{psib}) can be still summarized by denoting  
\begin{eqnarray}
\uppsi_g = \uppsi +\uppsi(\Uptheta)\,,
\label{psi11}
\end{eqnarray}
meaning that an extended multivector $\clb$ is a sum that consists of $\uppsi\in \sec\cle$ and an $g$-dependent multivector part $\uppsi(\Uptheta)\in\sec\cle$. Here, 
$\uppsi(\Uptheta)\in\sec\left(\Upomega^0(M)\oplus\Upomega^1(M)\oplus\Upomega^2(M)\right)$, and  
\begin{eqnarray}
\uppsi(\Uptheta) &=& \upbeta^{\mu\nu}\Uptheta_{\mu\nu}+ \upbeta^{\mu\nu\rho}(\Uptheta_{\mu\nu}\upgamma_\rho {\clt{-}}  
  \Uptheta_{\rho\mu}\upgamma_\nu + \Uptheta_{\nu\rho}\upgamma_\mu)
\nonumber\\&+&{{p(\Uptheta_{23}\upgamma_{0}\upgamma_{1} \!-\! \Uptheta_{13}\upgamma_{0}\upgamma_{2} \!+\! \Uptheta_{12}\upgamma_{0}\upgamma_{3} \!+\! \Uptheta_{03}\upgamma_{1}\upgamma_{2} \!-\! \Uptheta_{02}\upgamma_{1}\upgamma_{3} \!+\! \Uptheta_{01}\upgamma_{2}\upgamma_{3})}},  
\label{pisa}
\end{eqnarray} 

An algebraic $g$-spinor field is a multivector field in the minimal left ideal $(\CC\otimes\clb)\pu {f_g}$, which can be written as,
\beq
(\uppsi_g)\pu(f_g) &=& (\uppsi + \uppsi(\Uptheta))\pu (f+f(\Uptheta))\nonumber\\
&=& (\uppsi)\pu f + \uppsi(\Uptheta)\pu f + (\uppsi)\pu f(\Uptheta) + (\uppsi(\Uptheta))\pu f(\Uptheta),\label{psidis1}
\eeq
when Eqs.~(\ref{f11}) and (\ref{psi11}) are employed. The last term $(\uppsi(\Uptheta))\pu f(\Uptheta)$ runs with $\mathcal{O}(\Theta^2)$, consisting of two terms, each one involving $\mathcal{O}(\Theta)$. Therefore, it can be discarded in calculating $(\uppsi_g)\pu(f_g)$. It is possible to find terms up to $\mathcal{O}(\Theta^2)$ for all the other terms. To do so, one has to use Eqs. (\ref{cfl}) and (\ref{q1}), as well as the Clifford product definition in order to implement products as $(\gamma_\mu\gamma_\nu)\underset{g}{}\gamma_\rho=\gamma_\mu\JJ\gamma_\nu\underset{g}{}\gamma_\rho+\gamma_\mu\wedge(\gamma_\nu\underset{g}{}\gamma_{\rho})$, $\gamma_\mu\underset{g}{}(\gamma_\nu\gamma_\rho)=\gamma_\mu\underset{g}{}\gamma_\nu\JJ\gamma_\rho+(\gamma_\mu\underset{g}{}\gamma_\nu)\wedge\gamma_\rho$ and so on.

\section{Extended Spinor Field Classification}
\label{scle2}
One can address the correlation among spinor fields in the Minkowski spacetime and extended $g$-spinor fields in the nontrivial topology. The extended spinor field classification can be engendered by symbolically emulating the spinor field classes~\cite{lou2}. The first three classes regard regular exotic spinor fields and in them also $\mathbf{{\scalebox{0.9}{$\textbf{J}$}}}_g, \mathbf{{\scalebox{0.9}{$\textbf{K}$}}}_g, \mathbf{{\scalebox{0.9}{$\textbf{S}$}}}_g \neq 0$:

\begin{itemize}
\item[$1_g$)] $\upsigma_g\neq0,\;\;\; \upomega_g\neq0$.

\item[$2_g$)] $\upsigma_g\neq0,\;\;\; \upomega_g= 0$.\label{dirac1}

\item[$3_g$)] $\upsigma_g= 0, \;\;\;\upomega_g\neq0$.\label{dirac2}

\item[$4_g$)] $\upsigma_g= 0 = \upomega_g, \;\;\;\mathbf{{\scalebox{0.9}{$\textbf{K}$}}}_g\neq0,\;\;\; \mathbf{{\scalebox{0.9}{$\textbf{S}$}}}_g\neq0$.%
\label{tipo4}

\item[$5_g$)] $\upsigma_g= 0 = \upomega_g, \;\;\;\mathbf{{\scalebox{0.9}{$\textbf{K}$}}}_g= 0, \;\;\;\mathbf{{\scalebox{0.9}{$\textbf{S}$}}}_g\neq0$.%
\label{type-(5)1}

\item[$6_g$)] $\upsigma_g= 0 = \upomega_g, \;\;\; \mathbf{{\scalebox{0.9}{$\textbf{K}$}}}_g\neq0, \;\;\; \mathbf{{\scalebox{0.9}{$\textbf{S}$}}}_g = 0$.
\end{itemize}
We have seen that the algebraic spinor in Eq. (\ref{rep1}), which is a Clifford algebraic element in the left minimal ideal $(\CC\otimes \cl _{1,3})f$, is isomorphic to the  4-spinor in Eq. (\ref{dire}), which is a column spinor $\uppsi$ in $\mathbb{C}^{4}$.
The Dirac adjoint of the 4-spinor $\bar\uppsi = \uppsi^\dagger \gamma^0$ is well known to be a row matrix, given by 
\beq
\bar\uppsi = \uppsi^\dagger \gamma^0 = (\uppsi_1^*, \uppsi_2^*, -\uppsi_3^*, -\uppsi_4^*).
\eeq
However, for the square matrix spinor $[\uppsi]\in(\CC\otimes \cl _{1,3})f\simeq {\rm M}(4,\CC)f$ in (\ref{rep1}), its Dirac adjoint is the  square matrix
\beq\label{rep122}
\overline{[\uppsi]} \simeq  
\begin{bmatrix}
\uppsi_1^*& \uppsi_2^*& -\uppsi_3^*& -\uppsi_4^* \\ 
0 & 0 & 0 & 0 \\ 
0 & 0 & 0 & 0 \\ 
0 & 0 & 0 & 0%
\end{bmatrix}\in  {\rm M}(4,\CC)f\simeq(\CC\otimes \cl _{1,3})f,
\eeq 
which can implemented by $\overline{[\uppsi]} = \gamma_0[\uppsi]^\dagger \gamma_0^{-1}$, in the left minimal ideal representation ${\rm M}(4,\CC)f$, or by $ \widetilde{[\uppsi]^*}$ in $(\CC\otimes \cl _{1,3})f$, where the last equality involves taking the reversion of the complex conjugate of the Clifford algebraic object, in the isomorphic the left minimal ideal $\uppsi\in(\CC\otimes \cl _{1,3})f$ \cite{lou2}.

The scalar $g$-bilinear covariant $\upsigma_g$ can be straightforwardly calculated by taking the definition of the algebraic $g$-spinor field as a multivector field in the minimal left ideal $(\CC\otimes\clb)\pu {f_g}$, which can be written as in Eq. (\ref{psidis1}). Since we are taking the algebraic definition of the $g$-spinor field as an element in the minimal left ideal $(\CC\otimes\clb)\pu {f_g}$, therefore its Dirac adjoint  reads $\widetilde{(\uppsi_g^*)\pu(f_g^*)}$. Therefore, the scalar $g$-bilinear covariant reads:
\beq
\upsigma_g&=&{\widetilde{(\uppsi^*_g)\pu(f_g^*)}}\pu [(\uppsi_g)\pu(f_g)]\nonumber\\
&=& {\widetilde{(f_g^*)}\underset{\!\!\!\!\!\!\!\!\!\!\!\!\!\!\!\!{\scalebox{0.65}{$\emph{g}$}}}{\widetilde{(\uppsi^*_g)}}}\pu [(\uppsi_g)\pu(f_g)],
\eeq
where each term in the above equation can be read off either by Eqs. (\ref{f11}, \ref{spinorq}) or by taking directly Eq. (\ref{psidis1}). The other $g$-bilinear covariants can be analogously calculated using this method. In fact, 
\beq
{\bf J}_g&=&(J_{\mu})_g{\scalebox{0.8}{$\textsc{E}$}} ^{\mu },\eeq for 
\beq
(J_{\mu})_g =  {\widetilde{(f_g^*)}\underset{\!\!\!\!\!\!\!\!\!\!\!\!\!\!\!\!{\scalebox{0.65}{$\emph{g}$}}}{\widetilde{(\uppsi^*_g)}}}\pu (\upgamma_\mu)\pu [(\uppsi_g)\pu(f_g)].
\eeq
The 2-covector $g$-bilinear covariant reads
\beq
{\bf S}_g = (S_{\mu \nu })_g{\scalebox{0.8}{$\textsc{E}$}} ^{\mu }\wedge {\scalebox{0.8}{$\textsc{E}$}} ^{\nu },\eeq where  
\beq
(S_{\mu \nu })_g =  \frac{i}{2}{\widetilde{(f_g^*)}\underset{\!\!\!\!\!\!\!\!\!\!\!\!\!\!\!\!{\scalebox{0.65}{$\emph{g}$}}}{\widetilde{(\uppsi^*_g)}}}\underset{\!\!\!\!\!\!\!\!\!\!\!\!\!\!\!\!\!\!\!\!\!\!\!\!\!{\scalebox{0.65}{$\emph{g}$}}} {\left[{\upgamma _{\mu}},{\upgamma_{\nu}}\right]}\pu [(\uppsi_g)\pu(f_g)].
\eeq The pseudocovector $g$-bilinear covariant is given by
\beq
{\bf K}_g&=&(K_{\mu})_g{\scalebox{0.8}{$\textsc{E}$}} ^{\mu },\eeq for 
\beq
(K_{\mu})_g =  {\widetilde{(f_g^*)}\underset{\!\!\!\!\!\!\!\!\!\!\!\!\!\!\!\!{\scalebox{0.65}{$\emph{g}$}}}{\widetilde{(\uppsi^*_g)}}}\pu ({\upgamma_\mu}\pu {{\upgamma_5}_g})\pu [(\uppsi_g)\pu(f_g)],
\eeq
where ${(\upgamma_5)}_g = -i{\upgamma_0}\pu{\upgamma_1}\pu{\upgamma_2}\pu{\upgamma_3}$ 
and finally, the pseudoscalar reads 
\beq
\upomega_g =  -{\widetilde{(f_g^*)}\underset{\!\!\!\!\!\!\!\!\!\!\!\!\!\!\!\!{\scalebox{0.65}{$\emph{g}$}}}{\widetilde{(\uppsi^*_g)}}}\pu ({\upgamma_5}_g)\pu [(\uppsi_g)\pu(f_g)].
\eeq

Due to Eq. (\ref{exo}), one can express
\begin{subequations}
\beq
\upsigma_g &=& \upsigma + \upsigma(\Uptheta),\\
{\mathbf J}_g &=& {\mathbf J} + {\mathbf J}(\Uptheta),\\
{\mathbf S}_g &=& {\mathbf S} + {\mathbf S}(\Uptheta),\\
{\mathbf K}_g &=& {\mathbf K} + {\mathbf K}(\Uptheta),\\
\upomega_g &=& \upomega + \upomega(\Uptheta)\,. 
\eeq
\end{subequations}
With the $\Theta$ corrections, we can construct all the $g$-bilinear covariants. A complementary duality among the spinor fields in the standard classification and the extended spinor fields in the spinor field classes $1_g) - 6_g)$ can be now established.
\begin{itemize}
\item[$1_g$)] $\upsigma_g\neq0,\;\; \upomega_g\neq0$.
Since the scalar bilinear covariant $\upsigma_g$ is not equal to zero, and using the fact that $\upsigma_g = \upsigma + \upsigma(\Uptheta)$, new possibilities of combination emerge, depending on whether the scalar bilinear covariant 
$\upsigma$ equals or is not equal to zero, as well as the corresponding several possibilities for the pseudoscalar bilinear covariant $\upomega$. Therefore the following cases can be then derived:
\begin{enumerate}
\item[$i)$] $\upsigma = 0 = \upomega$. This case regards singular spinor fields in the standard spinor field classification and is compatible with the possibilities $\upsigma_g\neq0$ and $\upomega_g\neq0$, sufficing, for it, that $\upsigma(\Uptheta)\neq0$ and $\upomega(\Uptheta)\neq0$, respectively.
\item[$ii)$] $\upsigma = 0$ and $\upomega\neq 0$. This case regards usual spinor fields of type-3. 
The bound $\upsigma = 0$ can match the condition $\upsigma_g\neq0$. However, the constraint $\upomega\neq 0$ yields the additional bound 
$\upomega_g  \neq 0$ or, equivalently, $\upomega \neq- \upomega(\Uptheta)$. 
\item[$iii)$] $\upsigma \neq 0$ and $\upomega = 0$. This case matches the usual spinor fields of type-2. The constraint $\upomega = 0$ complies to $\upomega_g\neq0$. Nevertheless, since $\upsigma\neq 0$, the additional constraint $\upsigma_g  \neq 0$ has to be estipulated, corresponding to $\upsigma \neq-  \upsigma(\Uptheta)$. 
\item[$iv)$] $\upsigma \neq 0$ and $\upomega\neq 0$. This case regards usual spinor fields of type-1. 
In this case, the constraints $\upomega\neq -\upomega(\Uptheta)$ and $\upsigma \neq -\upsigma(\Uptheta)$ must be concomitantly required.
 \end{enumerate}
The conditions analyzed above have to be imposed for an extended $g$-spinor field to represent spinor fields in class $1_g)$.

\item[$2_g$)] $\upsigma_g\neq0,\;\;\; \upomega_g= 0$.\label{dirac1b}
Even though the constraint $\upsigma_g\neq0$ can be indeed compatible with both the equalities $\upsigma = 0$ and $\upsigma\neq 0$, the constraint $\upomega_g= 0$ yields $\upomega = - \upomega(\Uptheta)$, which does not necessarily vanishes. One must emphasize that the constraint $\upsigma\neq0$ can match the condition $\upsigma_g\neq0$ as long as $\upsigma\neq -\upsigma(\Uptheta)$. 
One can therefore summarize the results as follows:
\begin{enumerate}
\item[$i)$] $\upsigma = 0$, $\upomega\neq 0$. This case regards usual spinor fields of type-3. 
The constraint $\upsigma = 0$ can match the inequality $\upsigma_g\neq0$. However, since $\upomega\neq 0$, the ancillary constraint $\upsigma(\Uptheta)\neq0$, corresponding to $\upomega_g \neq0$, has to be foisted. Also, the condition $\upomega = -\upomega(\Uptheta)$ must hold, in such a way that $\upomega_g = 0$.
\item[$ii)$] $\upsigma \neq 0$, $\upomega\neq 0$. This case regards usual spinor fields of type-1. The constraints $\upomega = -\upomega(\Uptheta)$ and $\upsigma \neq - \upsigma(\Uptheta)$ have to be assumed.
 \end{enumerate}
\item[$3_g$)] $\upsigma_g= 0$, $\upomega_g\neq0$.\label{dirac2b}
In spite of the constraint $\upomega_g\neq0$ to match both the possibilities $\upomega = 0$ and $\upomega\neq 0$, the bound $\upsigma_g= 0$ yields the equality $\upsigma = - \upsigma(\Uptheta)$. One can sum up the following sub-possibilities:
\begin{enumerate}
\item[$i)$] $\upomega= 0$, $\upsigma\neq 0$. This case regards usual spinor fields of type-2. The constraint $\upomega = 0$ can be  satisfied by $\upomega_g\neq0$. Nevertheless, since $\upsigma\neq 0$, the ancillary constraint 
$\upsigma  = - \upsigma(\Uptheta)$ has to be required, as well as $\upomega(\Uptheta)\neq0$.
\item[$ii)$] $\upsigma \neq 0$ and $\upomega\neq 0$. This case regards usual spinor fields of type-1. The constraints $\upomega \neq -\upomega(\Uptheta)$ and $\upsigma = -\upsigma(\Uptheta)$ must be assumed.
 \end{enumerate}

\item[$4_g$)] $\upsigma_g= 0 = \upomega_g, \;\;\;\mathbf{{\scalebox{0.9}{$\textbf{K}$}}}_g\neq0,\;\;\; \mathbf{{\scalebox{0.9}{$\textbf{S}$}}}_g\neq0$.%
\label{tipo4b}

\item[$5_g$)] $\upsigma_g= 0 = \upomega_g, \;\;\;\mathbf{{\scalebox{0.9}{$\textbf{K}$}}}_g= 0, \;\;\;\mathbf{{\scalebox{0.9}{$\textbf{S}$}}}_g\neq0$.%
\label{type-(5)1b}

\item[$6_g$)] $\upsigma_g= 0 = \upomega_g, \;\;\; \mathbf{{\scalebox{0.9}{$\textbf{K}$}}}_g\neq0, \;\;\; \mathbf{{\scalebox{0.9}{$\textbf{S}$}}}_g = 0$.
\end{itemize}
Singular extended spinor fields $4_g$), $5_g$), and $6_g$) are similarly established by the constraint $\upsigma_g= 0 = \upomega_g$, yielding the conditions $\upsigma = -\upsigma(\Uptheta) \neq 0$ and $\upomega = -\upomega(\Uptheta) \neq 0$, which implies that the singular extended $g$-spinor fields correspond to regular, type-1, spinor fields. 
Table \ref{tata} summarizes the results obtained. 
 \medbreak
 \begin{table}
\centering
\begin{tabular}{||r|r?r|r||}
\hline\hline
\,&\,Exotic Spinor Fields\,&\,Spinor Fields\,&\,\\\specialrule{.1em}{.05em}{.05em} 
 type-$1_g$\,&\, $g$-regular\,&\,regular\,&\,type-1\\
 &\,&\,regular\,&\,type-2\\
 & \,&\,regular\,&\,type-3\\
& \,&\,flag-dipole\,&\,type-4\\
& \,&\,flag-pole \,&\,type-5\\
& \,&\,dipole \,&\,type-6\\\hline
type-$2_g$\,&\, $g$-regular\,&\,regular\,&\,type-3\\
 & \,&\,regular\,&\,type-1\\\hline
type-$3_g$\,&\, $g$-regular\,&\,regular\,&\,type-2\\
 & \,&\,regular\,&\,type-1\\\hline
 type-$4_g$\,&\, $g$-flag-dipole\,&\,regular\,&\,type-1\\\hline 
 type-$5_g$ \,&\, $g$-flag-pole\,&\,regular\,&\,type-1\\\hline
 type-$6_g$\,&\, $g$-dipole\,&\,regular\,&\,type-1\\\hline
  
\hline\hline
\end{tabular}\vspace{0.5mm}
\caption{\small{Correspondence among the classes of standard spinor fields and extended $g$-spinor fields, according to the spinor field classification.}}
\label{tata}
\end{table}

\section{Conclusions}
\label{scle3}

The classification encoded in Table \ref{tata} presents a straightforward way to probe the exotic structure that composes the bilinear form that endows the spacetime. Moreover, it reveals the duality between the standard and the exotic spinor fields, under the spinor field classification, by proposing which classes of spinor fields can be connected 
by the introduction of an additional exotic part in the spacetime metric. 
The usual case of spinor fields on 4-dimensional Lorentzian manifolds is recovered, as long as the $\theta$-function, which carries signatures of the nontrivial topology into the spin structure, is constant. This behavior is also inherited by both the extended spinor and Clifford bundles. Regarding the results presented in Table \ref{tata}, however, one should take additional care with this limit in the spinor level. The reason is that once a given extended spinor turns to the standard case, there is no clue coming from the formalism to which class the spinor goes. It is reasonable to assert that once the topological effects are to be disregarded, the extended spinor redounds on the class it belonged to before topological effects and extensions went in order. Even though this idea is sound from the physical point of view, this specific limit deserves more attention since it could lead to nontrivial Lounesto's class flipping, while preserving the spinorial degrees of freedom. Delving into this question, we hope to thoroughly answer it soon.  

\subsection*{Acknowledgements}
JMHS thanks the National Council for Scientific and Technological Development -- CNPq (grant No. 307641/2022-8) and Dr. Rodolfo J. Bueno Rog\'erio for very stimulating discussions. RdR thanks grants No. 2022/01734-7 and No. 2021/01089-1 of S\~ao Paulo Research Foundation (FAPESP), and grant CNPq No. 303390/2019-0. The authors are grateful to Deborah Fabri, for fruitful discussions and the figures.    

\end{document}